\documentclass[preprint,aps,floats,superscriptaddress,floatfix,onecolumn,amsmath,amssymb]{revtex4}
\usepackage{amssymb,amsmath}
\usepackage{amsmath,amssymb}
\usepackage{graphicx}
\usepackage{psfrag}
\usepackage{multirow}
\usepackage{color}
\usepackage{dcolumn}
\usepackage{bm}
\usepackage{color}
\usepackage{enumitem,kantlipsum}
\usepackage{bm}
\usepackage[normalem]{ulem}

\def\beq{\begin{equation}}
\def\eeq{\end{equation}}
\def\bea{\begin{eqnarray}}
\def\eea{\end{eqnarray}}
\begin{document}
\title{Smooth or shock: universality in closed inhomogeneous driven single file motions}
 \author{Tirthankar Banerjee}\email{shantoshisto38@gmail.com, tirthankar.banerjee@kuleuven.be}
 \affiliation{Instituut voor Theoretische Fysica, KU Leuven, 3001 Heverlee, Belgium}
 \affiliation{LPTMS, CNRS, Univ. Paris-Sud, Universit\'e Paris-Saclay, 91405 Orsay cedex, France}
\author{Abhik Basu}\email{abhik.123@gmail.com, abhik.basu@saha.ac.in}
\affiliation{Condensed Matter Physics Division, Saha Institute of
Nuclear Physics, Calcutta 700064, India}

\date{\today}
\begin{abstract}
We study the nonequilibrum steady states in a  unidirectional {or driven} single file motion (DSFM) of 
a 
collection of particles with   hard-core repulsion in a closed system. For driven propulsion that is {spatially} smoothly varying with a few discontinuities, we show that the steady states are 
broadly classified into two classes,  independent of any system detail:  (i) when the steady state current depends explicitly on the {conserved} number density $n$, and (ii) when it is independent of   $n$. 
This manifests  itself in the universal topology of the phase diagrams  {and fundamental diagrams (i.e., the current versus density curves) } for DSFM,  which are determined solely by the interplay between {two control parameters} $n$ and the minimum propulsion speed along the chain. 
 Our theory can be 
tested in laboratory experiments on 
driven particles in a closed  geometry.

\end{abstract}

\maketitle
 Directed single file motion (DSFM) implies unidirectional particle movement along narrow 
channels where the
particles cannot cross each other due to hardcore repulsion.
 It is  an inherently
 nonequilibrium process that consumes energy
for propulsion.  We are particularly interested in DSFM with spatially nonuniform propulsion and finite resources, i.e., fixed available number of particles. This {should} be relevant in wide-ranging systems, 
e.g., vehicular or pedestrian movement along closed network of roads with  bottlenecks having varying strength,  closed urban transport networks  with enforced speed variations~\cite{traffic,traffic2} and spatially varying electric fields 
in closed arrays of quantum dots~\cite{qdot}. {This study could form the basis for further research on weakly number conserving quasi one-dimensional (1D) transport models where particle number conservation approximately holds at time-scales shorter than any nonconserving processes; e.g., ribosome 
translocations along closed mRNA loops  with pause sites (for which ribosomes are typically re-initiated in translocation and breaking of ribosome number conservation is likely to affect only at relatively large time scales)~\cite{zia1,ribo,pause1}. It should also be useful in studies on the effects of quenched disorder on asymmetric exclusion processes with finite resources~\cite{zia}.}

  The general goal of this work is to theoretically understand the classes of steady states in spatially nonuniform systems with restricted one-dimensional (1D) motion with finite resources, and to elucidate their universal nature. For this,
we {construct a minimal} theory for DSFM with position-dependent propulsion speed and hardcore repulsion in
closed geometries,  with the total number of particles $N_{tot}$ 
 being conserved.  This theory adequately describes the interplay between inhomogeneity and conservation laws, and reveals the generic  universal nature of the nonequilibrium steady states.
It applies to  all {\em in-vitro} or {\em in-vivo} systems {where individual particles are non-active or weakly active, i.e., do not actively push or pull the neighbors strongly and} are  undergoing quasi-1D motion, without mutual passage and having number conservation. 
It can also be useful and serve as theoretical benchmark for quasi-1D systems with weak  particle non-conservation, e.g., binding factor mediated enhancement of the probability of loop formation in mRNA in eukaryotes~\cite{chou1}.
The results can be tested in carefully designed {\em in-vitro} experiments on the collective motion of driven particles along a 
nonuniform closed track.

 We focus on the steady state
densities in DSFM and their dependences on $N_{tot}$ and position-dependent propulsion.
In order to extract generic results from a minimal description without losing the essential physics,  we model DSFM by the 
well-known 1D totally asymmetric simple exclusion
process (TASEP), where each site can accommodate at most one particle that can 
hop only in one direction if the 
neighboring site is empty. TASEP with open boundaries is a simple 
model for nonequilibrium phase transitions in 1D open systems~\cite{derrida,krug-ori,zia1}.

In this article, we study closed TASEP with $N$ sites as a model for 
DSFM. Space-dependent propulsion is described by quenched hopping rates that are
spatially smoothly varying with finite number of discontinuities having single or multiple point minima. 
 The main results are:  (i) independent of the details of the heterogeneous hopping rates, there are generically two 
{classes of steady states} delineated by the steady state current $J$:  { (a)} { when} $J$ depends on mean density
$n=N_{tot}/N$ ($0<n<1)$ explicitly (hereafter {\em smooth phase}), and  { (b)} { when} $J$ is 
independent of $n$, characterized by { a} {phase separation with}  localized (LDW) or delocalized 
(DDW) domain walls (hereafter {\em shock phase}), (ii) the 
phases and the {\em reentrant} transitions between them are controlled by 
the interplay between $n$ and  the global minima $q_{min}$ of the position-dependent propulsion speed, (iii) moving shocks appear only for  multiple {\em global} minima in propulsion speed; 
multiple local minima with only one global minimum only produce a localized shock, and {(iv) while accumulation of particles where the hopping rate is low is na\"ively expected, we show below that the position of the peak of the density in the shock phase can actually be anywhere in the system, being controlled by $n$.}

\begin{figure}[htb]
\includegraphics[height=6.5cm]{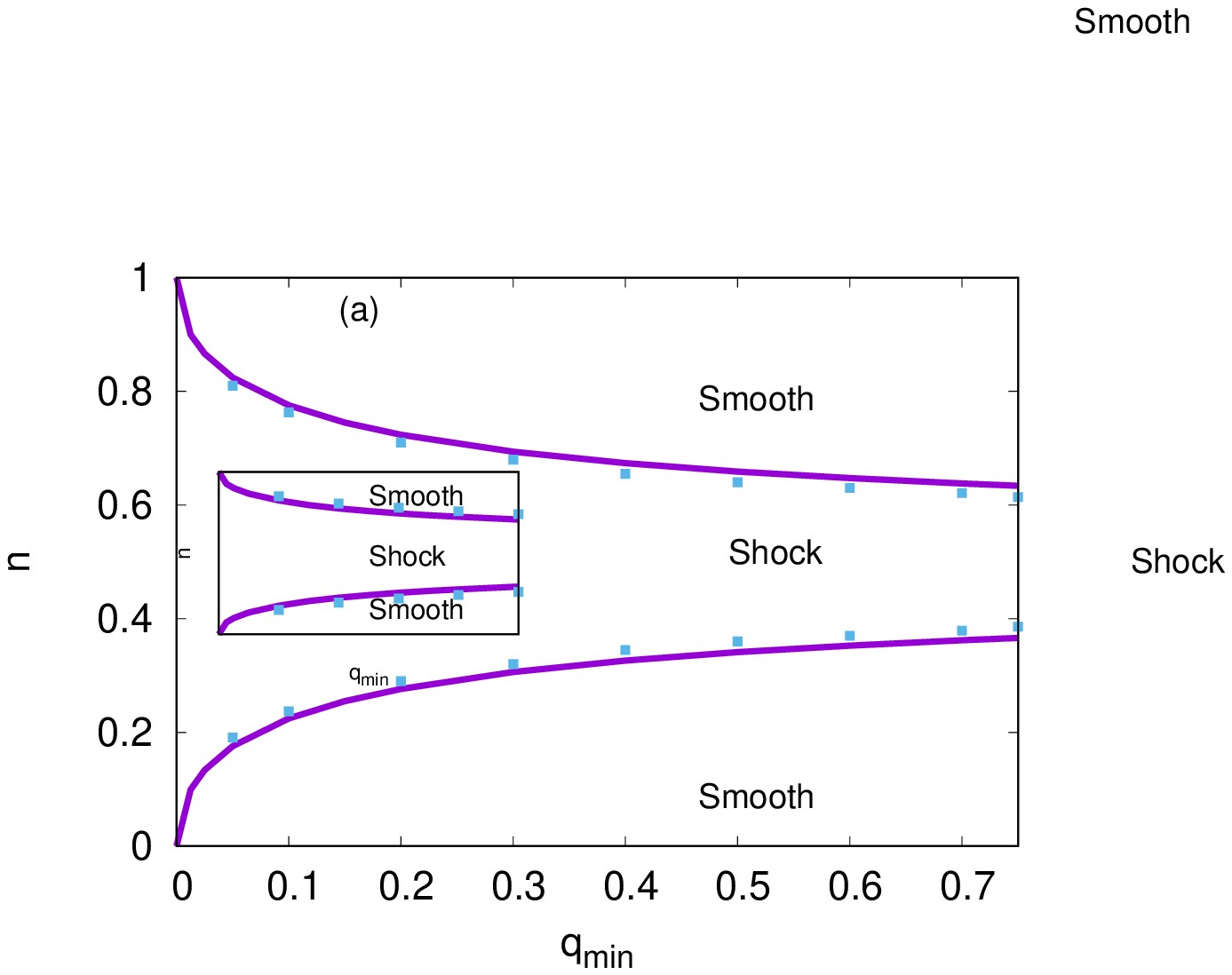}\\
\includegraphics[height=6.5cm]{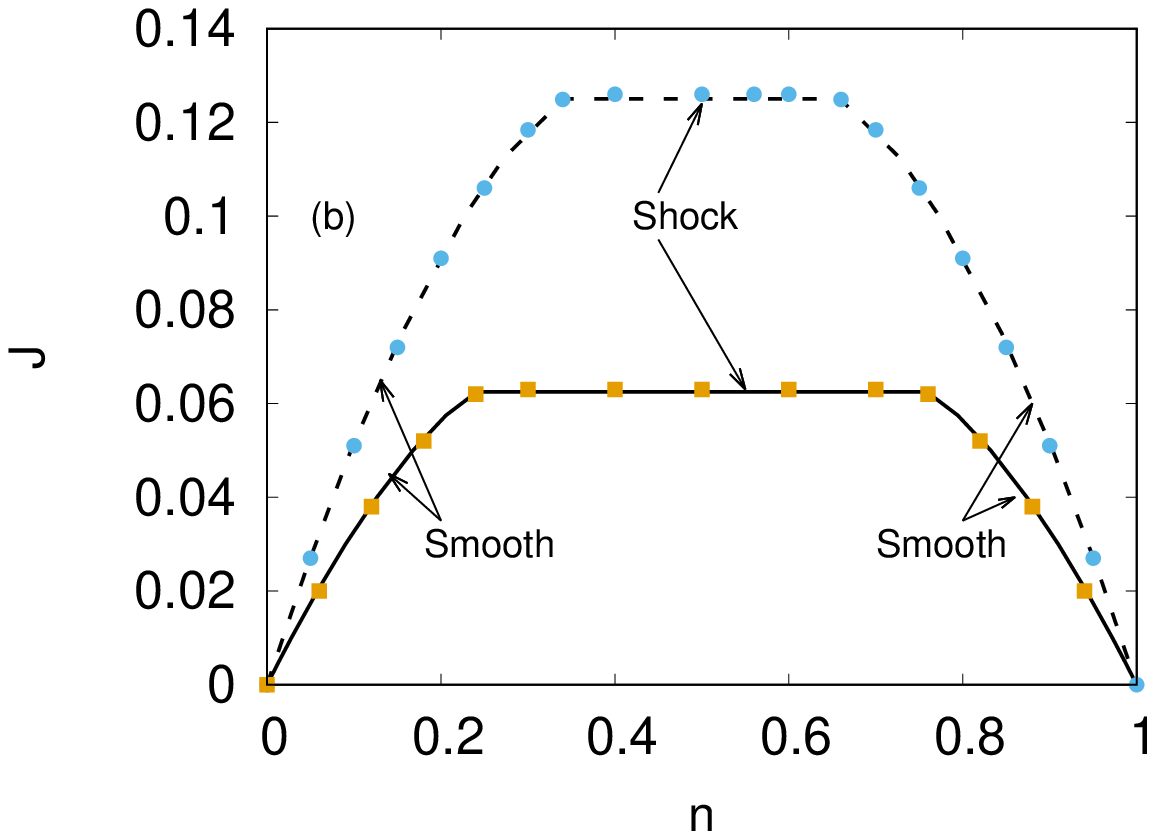}
\caption{(a) Illustrations of {\em universal topology} of the phase diagrams  in the $q_{min}-n$ plane: (i) $q(x)=(x-0.5)^2+q_{min}=q_1(x)$, and (ii) (inset) $q(x)=0.5x^2+q_{min}=q_2(x)$ (with $n=[0,1]$ and
$q_{min}=[0,0.75]$). The range of $q_{min}$ is chosen in such a way that $q(x)$ does not exceed unity anywhere.
Magenta lines and blue points are {overlapping} MFT and MCS results, respectively.
(b)  Fundamental diagrams ($J$ vs $n$) for $q_1(x)=(0.5-x)^2 +0.5$ (dashed curves) and $q_2(x)=0.5 x^2 + 0.25$ (solid curves) as chosen above for (a). 
The different phases that exist on the {fundamental diagrams} are marked.  Both curves have the same form (see text). {Note the saturation of current $J$ with respect to the density $n$ in the shock phase.}
Curves and points represent MFT and MCS results, respectively.}  \label{phase-diag}
\end{figure}

 This article shows how the general concept of universality, well-developed for equilibrium systems, applies  for {\em spatially-varying steady state density profiles in driven inhomogeneous systems with number conservation. This remains hitherto unexplored.} More specifically,
the topology of the phase diagrams plotted as functions of $n$ and $q_{min}$ { and {the} {associated fundamental diagrams} (i.e., the $J$ versus $n$ {plots}) } is  argued to be {\em universal}, independent of the 
{precise} hopping rate 
functions; see Fig.~\ref{phase-diag}.


 The model has a hopping rate $q_i\leq 1$ at a site $i$.  
The dynamics clearly 
conserves the total particle number $N_{tot}=\sum_i^N n_i$, where $n_i$ is the 
occupation of site $i$. 
The dynamics of TASEP is formally given by
rate equations for every site, {which are} not closed~\cite{tasep}. 
 In mean-field theory (MFT), we write down the dynamical equations for TASEP in closed forms, amenable to analytical treatments.
We label the sites by $x=i/N$; in the thermodynamic limit 
$N\rightarrow\infty$, $x$ effectively becomes a continuous variable confined
between $0$ and $1$. In this parametrization, the hopping rate function 
is given by $0< q(x)\leq 1$; we assume $q(x)$ to be piece-wise continuous, smooth, 
slowly varying functions of $x$ with a few point minima. 
Further, we define $\rho(x)=\langle n_i\rangle$ as 
the density at 
$x$; here $\langle...\rangle$ refers to temporal 
averages in the steady states. Studies on quenched heterogeneous TASEP has a long history; see, e.g., Refs.~\cite{mustansir,stinchcombe,others,shaw} for some studies on different aspects of heterogeneous TASEP. Our model complements these existing works, and primarily investigates the notion of universality not discussed elsewhere.



 In the steady states
\begin{equation}\label{mf-eq}
\frac{\partial\rho}{\partial t}=- 
\frac{\partial}{\partial x}[q(x)\rho(1-\rho)]=0,
\end{equation}
 in MFT over a range of $x$ in which $q(x)$ is smooth ~\cite{erwin-rev}. 
This  gives 
\begin{equation}
 q(x)\rho (x) [1-\rho (x)]=J,\label{mf-eq1}
\end{equation}
where $J$, a constant  (yet unknown) is the steady state
current. 
 Equation~(\ref{mf-eq1}) has two spatially nonuniform
solutions $\rho_+(x)$ and $\rho_-(x)$:
\begin{eqnarray}
 \rho_+(x)&=&\frac{1}{2}\left[1 + \sqrt{1-4 
J/q(x)}\right]>\frac{1}{2},\label{rho+}\\
 \rho_-(x)&=&\frac{1}{2}\left[1 - \sqrt{1-4 J/q(x)}\right]<\frac{1}{2},\label{rho-}
\end{eqnarray}
for all $x$. Clearly, both $\rho_+(x)$ and 
$\rho_-(x)$ are smooth functions for all $x$, except where $q(x)$ itself is 
discontinuous. 
  { Since $\rho(x)>0$} for all $x$,  $1-\frac{4 J}{q(x)} \geq 0$. 
{Thus}
\begin{equation}
 J \leq q(x)/4,
\end{equation}
  everywhere. 
The maximum  allowed value of $J$ is thus independent of $n$: 
\begin{equation}
 J_{ {max}} = q_{min}/4,\label{jmax}
\end{equation}
whence $\rho_+(x_0)=\rho_-(x_0)$; $x=x_0$ is the location of $q_{\rm min}$;
 see  
Ref.~\cite{krug-brazi} for analogous result in a disordered exclusion model.

 { We now 
outline the derivation of the results announced in the beginning of this article and the principles for the construction of Fig.~\ref{phase-diag}, 
followed by some illustrations of the phases for some {representative} $q(x)$. 
Since $\rho_+(x) (\rho_-(x)) > (<)1/2$ for all $x$ except at isolated points 
$x_0$ where $q(x)=q_{min}$, for a given $q(x)$   if $n$ is {\em sufficiently close} to 1/2, 
 $\rho(x)$ must be a 
combination of $\rho_+(x)(> \frac{1}{2})$ and $\rho_-(x)(< \frac{1}{2})$. This is the {\em shock phase} mentioned above.
In contrast, when $n$ approaches 0, there are only a few particles in the system 
and $\rho(x)=\rho_-(x)$ in the steady state throughout 
the system. Analogously, for $n$ approaching unity, $\rho(x)=\rho_+(x)$ in the steady state. 
The smooth phases with $\rho(x)=\rho_-(x)<1/2$, and $\rho(x)=\rho_+(x)>1/2$ are {\em spatially nonuniform}, and hence generalize the spatially {uniform} low density (LD) and high density (HD) phases, 
respectively, of TASEP {with open boundaries}~\cite{erwin-rev}. Thus, a {\em re-entrant smooth-shock-smooth} 
nonequilibrium phase transition is expected as $n$ rises from 0 to 1 for {any given} $q(x)$. The precise boundaries between these phases for a given $n$ and $q(x)$ - which will tell us for a given $q(x)$ how close $n$ must be to 1/2 for the smooth-shock transition - are obtained by imposing particle number conservation on (\ref{rho+}-\ref{rho-}) and using (\ref{jmax}); see below.

 Current $J$ is fixed by 
the particle number conservation:
\begin{equation}
 \int_0^1 \rho_a(x)=n,\;a=+\,-.\label{particle-cons1}
\end{equation}
  From (\ref{rho+}) and (\ref{rho-}),  the 
maximum (minimum) of $\rho_-(x)$ ($\rho_+(x)$) coincides with the minimum of 
$q(x)$, a fact borne out by 
our Monte-Carlo Simulation (MCS) studies: $q_{min}$ 
effectively acts as a bottleneck, and as a result, particles tend to accumulate 
behind it (see below).

As $n$ increases 
from zero, $J$ rises and eventually reaches 
$J_{max}$. For 
$J=J_{max}$, $\rho_+(x_0)=\rho_-(x_0)$, where $x_0$ is the location of 
$q_{min}$.  { On increasing $n$ further}, the additional particles are accommodated by 
representing $\rho(x)$ as a combination of $\rho_-(x)$ and 
$\rho_+(x)$ which meet smoothly at $x_0$. Since particles {should} accumulate 
behind the {\em bottleneck} at $x_0$, we expect that additional particles will go 
over to the high density solution represented by $\rho_+(x)$.  Since 
we have a closed system, the two solutions must meet at another point $x_w$, 
such that $\rho_+(x_w)>\rho_-(x_w)$ (since $\rho_+(x)=\rho_-(x)$ only at $x=x_0$), leading to a {\em discontinuous jump} in 
the form of a localized domain wall (LDW) in 
$\rho(x)$ at $x_w$,  thus giving rise to the {\em shock phase}, with a jump $\overline \rho$ {given by}
\begin{equation}
 \rho_+(x_w) -\rho_-(x_w)=\overline\rho,\label{jump}
\end{equation}
controlled by $n$ and the functional form of $q(x)$.
As more 
particles are added, $x_w$ shifts to make the region of existence for $\rho_+(x)$ larger and 
$\rho_-(x)$ smaller. This indeed leaves $J=J_{max}=q_{min}/4$ 
unchanged. { Thus the current in the shock phase saturates to its maximum value $J_{max}$.} This continues till $\rho_+(x)$ spans the full system. 
 Thus, as 
$n$ rises from the low to moderate values, a smooth-to-shock transition is 
encountered.  Interestingly, independent of the form of $q(x)$ this 
transition is {\em reentrant} - since, as $n$ rises further, the system moves from shock phase
to smooth phase again, { with $\rho_+$ now being the only valid solution}. { This reentrant transition can also be understood from the particle-hole symmetry of the model. Since particle density $\rho_+(x)$ can be interpreted as the hole density $1-\rho_+(x)=\rho_-(x)$, if $\rho_-(x)$ is a steady state solution for overall particle density $n$, $\rho_+(x)$ is a steady state solution for particle density $1-n$.} This picture remains valid even when there are additional local minima (but only one global 
minimum $q_{min}$ at $x_0$):  $J_{max}$ is still controlled solely by $q_{min}$ [Eq.~(\ref{jmax}) above], {with $\rho_+(x_0)=\rho_-(x_0)$;  the other local minima having no effect on $J_{max}$ are effectively screened in any steady state current measurements in shock phase}. However,
the 
form of the LDW, {i.e., the functional form of $\rho(x)$,} depends on any local minima through its 
dependence on the full form of $q(x)$.}

{Assuming only one global minimum for $q(x)$,
at the phase boundary between smooth and shock phases, 
$J=J_{max}=q_{min}/4$ and $\rho(x)=\rho_-(x)$ (for $n<1/2$) for all $x$, or  
for $n>1/2$,  $\rho_+(x)$ for all $x$. Thus
\begin{eqnarray}
 \int_0^1 dx\rho_\pm(x, J_{max}) &=& 
\int_0^1 dx \frac{1}{2}[1\pm\sqrt{1-\frac{q_{min}}{q(x)}}] = n,
\label{phase1}
\end{eqnarray}
give the quantitative dependence of $n$ on $q_{min}$ for the re-entrant transition, or equivalently, the boundaries between smooth and shock phases in Fig.~\ref{phase-diag} (a).  As our arguments above are independent of the precise form of $q(x)$, 
the topology of the phase diagrams in Fig.~\ref{phase-diag} (top) should remain independent of the 
precise forms of $q(x)$ having same $q_{min}$. This is the  universality in DSFM mentioned in the beginning that is also manifest in the {fundamental diagrams} in Fig.~\ref{phase-diag} (bottom), obtained from (\ref{mf-eq1}) and (\ref{particle-cons1}). The results in Fig.~\ref{phase-diag} are also obtained from the MCS studies, that corroborate the MFT predictions closely. This holds true even if there are multiple global minima with value $q_{min}$; 
in this case, shock phases   correspond to moving shocks (see below).  } {Notice that the topology of the phase diagram and the corresponding fundamental diagram are {\em same} as those obtained in Ref.~\cite{lebo}, where a single slow site controls the current, establishing an equivalence between the model  of Ref.~\cite{lebo} with a single slow site and our model here. The strength of the single slow site in the model of Ref.~\cite{lebo} corresponds to $q_{min}$ here.}

\begin{figure}[htb]
\includegraphics[width=8.3cm]{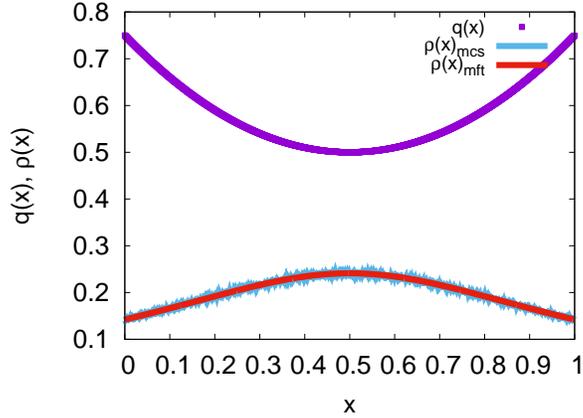}\hfill
\caption{Plot of  $\rho(x)$ versus $x$ in the LD phase 
($n=0.2$). Magenta line indicates the hopping rate function $q(x)=(x-0.5)^2+0.5$. 
Red continuous line and blue points denote $\rho_-(x)$, respectively, from MFT and   MCS studies (see text). Excellent agreement between 
MFT and MCS results is clearly visible. We have used $N=2000$.}  \label{ld-qtas}
\end{figure}

\begin{figure}[htb]
\includegraphics[width=8.3cm]{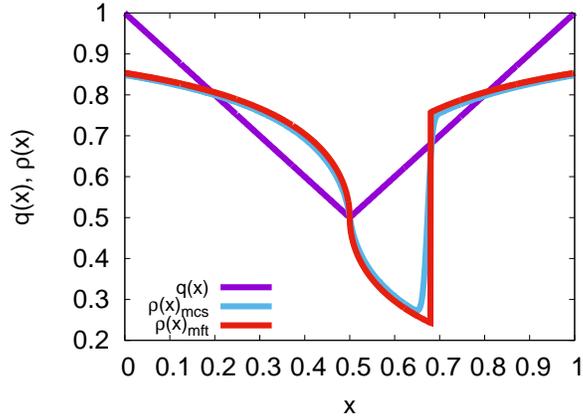}\hfill
\caption{Plot of $\rho(x)$ versus $x$ in the shock phase with $n=0.7, N=2000$. Magenta line indicates $q(x)=1-x, 0\leq x<0.5$ 
and $q(x)=x, 0.5\leq x<1$. Red line and blue points, respectively, represent MFT and MCS results for $\rho(x)$.}
 \label{cont-min1}
\end{figure}

 We now illustrate the phases with few representative choices for $q(x)$. See Fig.~\ref{ld-qtas} for a plot of $\rho(x)$ versus $x$ in the smooth phase; see also Figs.~(\ref{hd-club-app}-\ref{ap6}) 
in Appendix for plots of $\rho(x)$ with different $q(x)$ in the smooth phase; good agreements between MFT and MCS results are evident.  
Further  see Fig.~\ref{cont-min1} for a plot of the density in the shock phase  for a choice of $q(x)$, again showing strong agreements between MCS and MFT results;
see also Fig.~\ref{ldw-club-app} in Appendix. 

{ Notice that the peak of the density in the smooth phase in Fig.~\ref{ld-qtas} coincides with the location of $q_{min}$. As argued above, this can be understood from the form of $\rho_-(x)$ as given in (\ref{rho-}) - clearly, the maximum of $\rho_-(x)$ must coincide with the minimum of $q(x)$.  In contrast, the location of the extrema of the density profiles in the shock phase have no such relation with the minimum of $q(x)$. Instead, as we note in Fig.~\ref{cont-min1}, $\rho(x)$ is  continuous at the location $x_0$ of $q_{min}$, so long as $q(x)$ itself is continuous at $x_0$ (since $\rho_+=\rho_-$ at $x_0$). If however, $q(x)$ itself is discontinuous at $x_0$, the density is also discontinuous there; see Fig.~\ref{ldw-club-app} (right) in Appendix. In both these cases, the density has a discontinuity {\em elsewhere}  whose location is controlled by $n$ for a  given $q(x)$ with $q(x)$ being {\em continuous} there. This is one of the principal results in this article. Furthermore, given that $\rho_+(x)$ reduces and $\rho_-(x)$ grows when $q(x)$ grows in $x$ [see Eqs.~(\ref{rho+}) and (\ref{rho-}) above], the density in the shock phase being a linear combination of $\rho_+(x)$ and $\rho_-(x)$, can have a peak anywhere in the system that is controlled by $n$. }

\begin{figure}[htb]
\includegraphics[width=8.3cm]{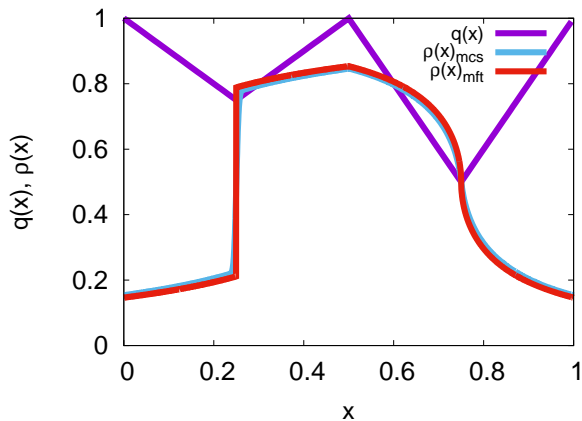}\hfill
\caption{One LDW with $n=0.6, N=2000$ and $q(x) = 1-x$, $x = [0, 0.25]$; $q(x) = 0.5 + x$, $x = [0.25, 0.5]$; 
$q(x) = 2-2x$, $x = [0.5, 0.75]$; $q(x) = 2x-1$, $x = [0.75, 1]$ with two 
local minima  of $q(x)$ at x = 0.25 and x = 0.75. Excellent agreement between MFT and MCS is observed.}  \label{discont-gro}
\end{figure}

 We further illustrate  the occurrence of a single LDW in the shock phase with $q(x)$ having multiple local but one global minimum  in Fig.~\ref{discont-gro}, with $\rho_+(x) = \rho_-(x)$ only at the location of the global minimum of $q(x)$, in agreement with the theoretical predictions. 

\begin{figure}[htb]
\includegraphics[width=8.3cm]{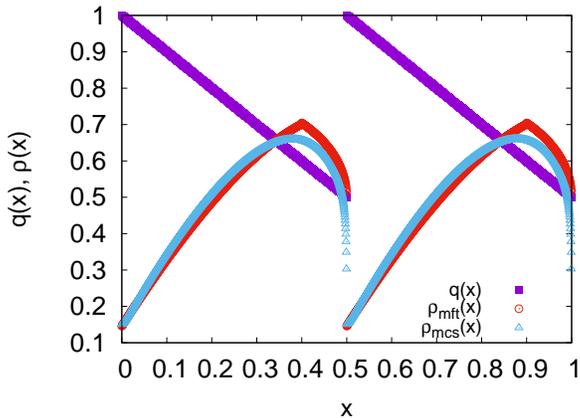}\hfill
\caption{Plot of $\rho(x)$ versus $x$ showing two DDWs; $n=0.5$. Magenta lines indicate $q(x)=1-x, 0\leq x<0.5$ and $q(x)=1.5-x, 0.5\leq x<1$; red line and blue points, respectively, are MFT and MCS results for DDW, which agree with each other. }  \label{ddw-2000}
\end{figure}

When there are multiple global minima,   the above-mentioned theory of LDW breaks down and very different physics emerges in the shock phase. For instance, consider 
 $q(x)$ to have only two global minima at  diametrically opposite points $x_1$ and 
$x_2$: $q(x_1)=q(x_2)=q_{min}$,  see Fig.~\ref{ddw-2000} for such a choice of $q(x)$. Thus there are now two effective bottlenecks at 
$x_1$ and $x_2$ which split the ring into two identical segments, say $T_A$ and $T_B$~\cite{frey,niladri1}, 
of { equal} length. Furthermore, from (\ref{rho+}) and (\ref{rho-}) we have
\begin{equation}
 \rho_-(x_1)=\rho_+(x_1)=\rho_-(x_2)=\rho_+(x_2).\label{minsrho}
\end{equation}
In each of $T_A$ and $T_B$,  MFT  for the shock phase
applies. Therefore, two domain walls are 
expected. Particle number conservation then  can only  yield  a single relation  between the two domain wall positions and $n$, and cannot determine both  $x_{w1}$ 
and $x_{w2}$  separately.  This means that a shift in $x_{w1}$ can be balanced by an 
equivalent reverse shift in $x_{w2}$ that still satisfies particle number conservation.  Due to the inherent stochasticity of the system, all 
possible solutions of $x_{w1}$ and $x_{w2}$ that are consistent with particle number conservation are visited by the system, if waited 
long enough. This leads to  {\em two} DDWs. Under long time 
averages, envelops of the moving DDWs will be observed;  see Fig.~\ref{ddw-2000} and Appendix for detailed calculations.

 As the DDWs in Fig.~\ref{ddw-2000} move, $T_A$ and $T_B$ exchange particles. 
Particle number conservation ensures that when a particle enters (leaves) $T_A$, it must necessarily leave (enter) $T_B$, 
leading to perfect synchronization of the DDWs in $T_A$ and $T_B$; see Appendix for additional discussions 
 and an associated kymograph in Fig.~\ref{kymo2}.

If there are more than two global minima of $q(x)$, then there will be as many  sub-channels ($T_A,T_B,T_C...$), and thus, as many DDWs.  However, the argument for synchronization breaks down in such cases; thus
 there should be no synchrony in movement between any two of the DDWs; 
see Fig.~\ref{ddw-club-app} (left) (and the corresponding kymograph Fig.~\ref{ddw-club-app} (right)) and related discussions in Appendix.



{The shocks in open TASEPs, which are {\em always} delocalized, appear when the entry ($\alpha$) and exit ($\beta$) are equal and not exceeding 1/2. This is equivalent to $J_{in}=J_{out}$, where $J_{in}$ and $J_{out}$ are, respectively, the currents determined by the entry side and exit side boundaries. The shock phase of the present model is analogous to  the shocks in open TASEPs. To see this we note that our ring model can be viewed as an open TASEP with effective entry ($\alpha_e=\rho_-(x_0)$) and exit ($\beta_e=1-\rho_+(x_0)$) rates that is joined at $x_0$, the location of $q_{min}$ (assuming a single global minimum)~\cite{niladri1,frey,tirtha1}. From the condition of the shock phase, $J_{in}=J_{out}=J_{max}=q_{min}/4$ holds automatically. That we obtain an LDW here as opposed to a DDW in an open TASEP under equivalent conditions is due to the strict particle number conservation here; see Refs.~\cite{lebo,niladri1,tirtha1,zia}. When there are two or more global minima of $q(x)$, the system can be thought to consist of those many TASEPs joined together to form a ring; see Refs.~\cite{niladri1,tirtha1}. The conditions for the shock phase is satisfied in all these TASEP segments simultaneously, leading to formation of a domain wall in each of them. As explained in Refs.~\cite{niladri1,tirtha1}, in this situation number conservation cannot pin the domain walls completely giving rise to DDWs; see also Appendix. We lastly note one point of dissimilarity between the shocks here and the DDWs in an open TASEP: for the latter, the current is less than its maximum value, where as in the shock phase here, the current  necessarily saturates at its maximum value of $J_{max}$. }



We have thus developed the theory for DSFM in a closed system with position-dependent propulsion by studying a closed TASEP with quenched hopping 
rates $q(x)$. This theory reveals the universal form of the phase diagrams and the fundamental diagrams of the model, for generic smooth $q(x)$  with a finite number of discontinuities and global minimum, but independent of the precise forms for $q(x)$. 
 Our theory is
sufficiently general and   applies to any smoothly varying $q(x)$ with finite 
number of discontinuities; it generalizes the analyses in Ref.~\cite{stinchcombe}.
From the perspectives of nonequilibrium systems, these results generalize the studies in 
Refs.~\cite{lebo,mustansir,stinchcombe1,niladri1,tirtha1}.  In the more complex situation where individual particles  can {\em actively} push (or pull) its neighbor,  thus  only dynamically modulating the effective hopping rate, we expect our results to remain valid for weak activity. For strong activity, competition between background heterogeneous hopping and the active processes will determine the steady state, whose full analysis is beyond the scope of the present work. { We have here studied smooth $q(x)$ with a small number of discontinuities. There are {\em in-vivo} situations where $q(x)$ could be rapidly fluctuating, e.g., the DNA strands~\cite{dna}. It would be interesting to study to what degree our results remain valid for rapidly fluctuating $q(x)$.} 

 Lastly, MFT developed here neglects spatial density correlations, which is an approximation. This may be improved by systematically including two-point correlations; see, e.g., Ref.~\cite{shaw}. It has been shown that extended particles instead of point particles significantly affect the steady state densities of an open TASEP with site-dependent hopping rate; 
see Refs.~\cite{shaw,dan}. 
It will be interesting to  study this effect in a closed TASEP.  

 This theory may be verified in model experiments on the 
collective motion of {driven} particles with light-induced activity~\cite{bechinger-jphys}  in a closed narrow circular 
channel~\cite{bechinger,bechinger-rmp}. Unidirectionality of the motion can be ensured by suppressing 
rotational diffusion, e.g., by choosing ellipsoidal particles with the channel width shorter than the long axis of the particle everywhere, or by using dimer particles. Propulsion speed can be tuned by applying patterned  or spatially varying
illumination~\cite{bechinger-jphys}. Steady state densities can be measured by microscopy with image processing. While technical challenges are anticipated in setting up appropriate experimental arrangements, we hope this  will be realized in near future. 



{\em Acknowledgement:-} We thank M. Khan for constructive suggestions and careful reading of the manuscript.
We also thank C. Maes for his critical comments on the manuscript.
The authors gratefully acknowledge  partial 
financial 
support from the Alexander von 
Humboldt Stiftung, Germany under the Research Group 
Linkage Programme (2016).

\appendix

\section{Density profiles for delocalized domain walls}

We calculate here  the steady state density profiles when $q(x)$ has two symmetrically placed global minima of same value. The system then can be considered to
 consist of two TASEP chains of equal size~\cite{niladri1}, say $T_A$ (with $0\leq x\leq 1/2$) and $T_B$ (with $1/2\leq x\leq 1$), each spanning from one global minimum of $q(x)$ to the other.
While the total particle number in the ring is conserved, the number of particles in each of $T_A$ and $T_B$ can fluctuate. { We closely follow Ref.~\cite{erwin-tobias} in our analysis below.}
\par
Now consider one delocalized domain wall (DDW) in each of $T_A$ and $T_B$. Let $x_w^A$ and $x_w^B$ be the 
instantaneous positions of the DDWs in $T_A$ and $T_B$, respectively and the respective heights be  $\Delta_A(x_w^A)$ and $\Delta_B(x_w^B)$. 
We note here that the DDW heights are explicit functions of their positions, since the steady state density is not uniform for an arbitrary $q(x)$.
 
 Now, increasing the number of particles in $T_A$ by $1$ would imply shifting $x_w^A$ by an amount $\delta x_w^A=\frac{-1}{L\Delta_A}$. 
Similarly, decrease of a particle would mean $\delta x_w^A=\frac{1}{L\Delta_A}$. In order to understand why this is so, we note that the 
'height $\Delta_A$ of the domain wall (DW) at $x_w^A$' means that $\Delta_A$ number of excess particles are needed to fill up one lattice spacing ($=\frac{1}{L}$), 
or to cause one lattice spacing leftward/rightward movement of the DW (and thus, the above values of $\delta x_w^A$). Let us now note that there are two
basic processes which can alter the number of particles individually in $T_A$ and $T_B$, i.e., if a particle enters $T_A$ through its left 
boundary (equivalent to a particle leaving $T_B$ through its right boundary) and vice-versa.
\par
For the following analysis, we will focus on $T_A$. Let $P(x_w^A,t)$ be the probability of finding a DW at $x_w^A$ at time $t$. 
For a given $x_w^A$, one can evaluate $x_w^B$ at time $t$, uniquely, using total particle number conservation.
The transition rate for a particle entering $T_A$ through the left boundary can be written as, $W_L=J_{in}= q(x=0)\alpha_e^A(1-\alpha_e^A), \delta x_w^A=\frac{-1}{L \Delta_A}$. Similarly, the transition rate for the particle leaving through the right boundary is given by $W_R=J_{out}=q(x=1/2) \beta_e^A(1-\beta_e^A), \delta x_w^A=\frac{1}{L \Delta_A}$. Here, $\alpha_e$ are $\beta_e$ are the densities at $x=0$ and $x=1/2$, respectively, in $T_A$, e.g., $\alpha_e=\rho(x=0)$ etc.

With these transition rates, we can calculate the average shift or the expectation value of the change, $\langle\delta x_w^A \rangle$, which is given by
the product of the increment (with sign) and the sum of the different transition rates:

\begin{equation}
\left <\delta x_w^A \right> = \frac{1}{L \Delta_A(x_w^A)}\left[J_{out}- J_{in}\right].
\end{equation}

It should be noted here, that the domain wall itself performs, a random walk about its mean position, $\overline{x_w^A}$. For the fixed point of the random walk, i.e., the value of $x_w^A$ for which $\langle\delta x_w^A \rangle=0$, we obtain,

\begin{equation}\label{dw_cond}
J_{out}-J_{in}=0,
\end{equation}
 the well known condition for formation of domain walls. {  In order to calculate the steady state profiles of the DDWs, we need to study the fluctuations in the DW positions that we do below.}

Using the expressions for the transition rates defined  above, we can write down the Master equation for $P(x_w^A,t)$, the probability of finding the DW at $x_w^A$ at time $t$.
\begin{eqnarray}
\frac{d P(x_w^A,t)}{d t} &=& \Sigma_{\delta x_w^A}[P(x_w^A+\delta x_w^A,t)
W(x_w^A+\delta x_w^A \rightarrow x_w^A)\nonumber \\ &-& P(x_w^A,t)W(x_w^A\rightarrow x_w^A + \delta x_w^A) ]
\end{eqnarray}
To proceed further, we employ Kramers-Moyal expansion~\cite{kramers} of the Master equation above around $\overline{x_w^A}$,
up to second order in $\delta x_w^A$. This gives,
\begin{equation}
\frac{d P(x_w^A,t)}{d t}=-\frac{\partial}{\partial y}\left[a(y) P(y,t)\right]+\frac{1}{2}\frac{\partial^2}{\partial y^2}\left[b(y) P(y,t)\right],
\end{equation}
where, $y=\delta x_w^A$, $a(y)=\Sigma_{y} y W (x_w^A+\delta x_w^A \rightarrow x_w^A)$ and $b(y)= \Sigma_{y} y^2 W (x_w^A+\delta x_w^A \rightarrow x_w^A)$.
Using the already known values for $W$ and $\delta x_w^A$, and Eq.~(\ref{dw_cond}) we arrive at the following results for $a$ and $b$:
\begin{equation}
a(x_w^A)=\frac{1}{L \Delta_A(x_w^A)}\left[-\alpha_e^A(1-\alpha_e^A)+\beta_e^A(1-\beta_e^A)\right]=0
\end{equation}
and
\begin{equation}
b(x_w^A)=\frac{1}{L^2 \Delta_A^2(x_w^A)}\left[\alpha_e^A(1-\alpha_e^A)+\beta_e^A(1-\beta_e^A)\right]>0
\end{equation}
Thus up to this order
\begin{equation}
\frac{d P(x,t)}{d t}=\frac{1}{2}\frac{\partial^2}{\partial x^2}\left[b(x)P(x)\right].
\end{equation}
Since $J_{in}=J_{out}$, the DW position effectively follows detailed balance condition. This means the fluctuations in the DW position should follow an equilibrium distribution in the steady state. 
Hence, the probability current, given by $J_{DW}(x)=\frac{\partial}{\partial x}\left[b(x)P(x)\right]=0$. This yields
\begin{equation}
P(x)=\frac{C}{b(x)},\label{prob-kramers}
\end{equation}
where $C$ is a constant which can be evaluated by the normalization condition on $P(x)$.

\subsection{Construction of the density profiles}

We can now construct the density profile $\rho(x)$ with the knowledge about $P(x)$. {Since the long time averaged steady state density involves averaging over $P(x)$}, we argue that 
\begin{equation}\label{density-prob}
\frac{\partial \rho}{ \partial x}= A P(x),
\end{equation}
where $A$ is a constant of proportionality.
Clearly from Eq.~(\ref{density-prob}), we can see if $P(x)={\rm const.}$, as in the case for a DDW in an open TASEP, $\rho(x)$ varies linearly with $x$, a known result. The constant $A$ in this example can be evaluated by the boundary conditions. In yet another example, for an LDW as $P(x) \propto \delta(x-x_w)$, $\rho(x)$ is a heaviside $\Theta$-function according to Eq.~(\ref{density-prob}), whose height can be determined using the boundary conditions. We now obtain the DDW steady state density profiles.

By using Eq.~(\ref{density-prob}) we write
\begin{equation}
\rho(x) = \tilde A \int \frac{dx}{b(x)}+D = A_1 \int dx (\rho_+(x) - \rho_-(x))^2+D,
\end{equation}
where, $\tilde A, A_1$ and $D$ are constants, and we have substituted the value of the DW height $\Delta_A(x)=\rho_+(x) - \rho_-(x)$ in $b(x)$. 
Using already derived expressions for $\rho_-(x)$ and $\rho_+(x)$ above, we finally arrive at the following expression for the DDW profile,
\begin{equation}
\rho(x) = A_1 \int dx(1-\frac{q_{min}}{q(x)})+D.
\end{equation}
The value of the constants can be fixed using the boundary conditions on $\rho(x)$. But there is one more undetermined quantity that we are yet to address.
The DDW in general has certain extent of wandering in $T_A$ { that is less than the length of the channel}, depending on the number density. We can have two situations: one, where $\rho(x)$ shows a
mix of LD and DDW profiles, or  where it is a mix of DDW and HD profiles. Therefore, within $x = [0,1/2]$, we can either have a situation with an LD phase from $x=0$ to say, $x=\overline{x}$,
followed by the DDW from $x=\overline{x}$ to $x=1/2$; or the situation with the DDW from $x=0$ to $x=\overline{x}$, followed by an HD phase from $x=\overline{x}$ to $x=1/2$.
Now, what determines the value of $\overline{x}$ is the condition $\int_0^{1/2} \rho_A(x) dx = n$, where $\rho_A(x)$ is the complete density profile for $T_A$ and $n$ is the number density (notice that $T_A$ and $T_B$ are identical, and both must have the same average number density).
If the DDW does not span the entire $T_A$, an additional unknown parameter $\overline x$ must be determined, which we fix numerically by using the particle number conservation. This in turn yields the complete density
profile. We use this scheme to obtain $\rho(x)$ for
 $q(x)=1-x$ for $x=[0,0.5]$ and $q(x)=1.5-x$ for $x=[0.5,1]$; see Fig.~\ref{ddw-2000} in the main text. Good agreement with the MCS result is clearly visible, establishing our analytical framework. 
 {
 \section{Synchronization of DDW movement}
 
 We now show that the two DDWs formed when $q(x)$ has two global minimal placed at diametrically opposite points 
 in the ring move with perfect synchrony. To do that we first consider the two basic microscopic processes in the dynamics that lead to the movement of individual domain walls. 
 \begin{itemize}
  \item (i) A particle leaves $T_A$ and enters into $T_B$. 
  \item (ii) A particle leaves $T_B$ and enters into $T_A$.
 \end{itemize}
Let $\delta x_w^A$ and $\delta x_w^B$ be the shifts in the instantaneous positions 
$x_w^A$ and $x_w^B$ of the domain walls in $T_A$ and $T_B$, respectively, due to the processes mentioned above.
Jumps in the densities at $x_w^A$ and $x_w^B$ are given by
\begin{eqnarray}
 \rho_+(x_w^A) -\rho_-(x_w^A)=\overline\rho_A,
  \rho_+(x_w^B) -\rho_-(x_w^B)=\overline\rho_B.
\end{eqnarray}
In general, $\overline\rho_A\neq \overline\rho_B$.

Since $\alpha_e = \beta_e$ for both $T_A$ and $T_B$, both processes (i) and (ii) take place with 
equal rate $W=\alpha_e(1-\alpha_e)$. 
By process (i) above, $\delta x_w^A= -WL^{-1}\overline\rho_A^{-1},\,\delta x_w^B = WL^{-1}\overline \rho_B^{-1}$. 
Similarly, by process (ii) above,
$\delta x_w^A= WL^{-1}\overline\rho_A^{-1},\,\delta x_w^B = -WL^{-1}\overline \rho_B^{-1}$.
Thus, 
\begin{equation}
 \langle \delta x_w^A\rangle +   \langle \delta x_w^B\rangle = 0 \label{sync1}
\end{equation}
identically. This means $\langle \delta x_w^A\rangle = -\langle \delta x_w^B\rangle$, 
which is the essence of synchronization of the movements of the two DDWs in $T_A$ and $T_B$. 
This synchronization manifests pictorially in a kymograph given in Fig.~\ref{kymo2}.

\begin{figure}[htb]
\includegraphics[width=9cm]{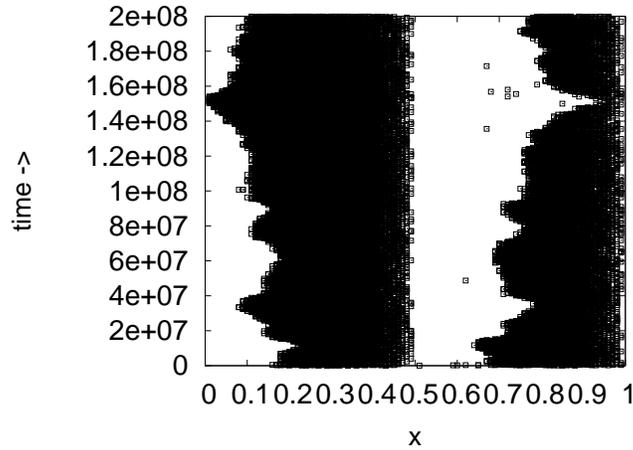}\hfill
\caption{Kymograph for $n=0.5, N=800$, $q(x)=1-x$ for $0\leq x  < 0.5;  q(x) =1.5-x$ for $0.5 \leq x<1$. See ~\cite{tirtha1} for technical details of obtaining kymographs of the same nature in closed TASEP systems.}  \label{kymo2}
\end{figure}

The above argument for synchronization breaks down when the number of global minima of $q(x)$,
which is same as the number of TASEP segments that make up the ring system, exceeds two. 
Now imagine $q(x)$ to have $N>2$ global minima of value $q_{min}$, placed at equal spacing. 
Thus the ring system can now be considered to be composed of $N$ TASEPs. 
The microscopic dynamical process for each TASEP consists of (i) receiving a particle from the previous 
TASEP and releasing a particle to the next one. Clearly, the general analog of (\ref{sync1}) would be
\begin{equation}
 \langle \delta x_w^A \rangle + \langle \delta x_w^B\rangle + \langle \delta x_w^C\rangle + ..=0.
\end{equation}
This ensures that
for any two successive 
TASEP segments the sum of the shifts of the corresponding DDW is no longer zero, leading to the loss of synchronization in the movement of the DDWs in any two
successive TASEP segments. 
}
 \widetext
 
\section{Density profiles}

We show below some representative plots of $\rho(x)$ versus $x$  from MCS studies along with MFT predictions. MCS studies have generally been performed with $N=2000$ with random sequential updates (except for Fig.~\ref{ddw-2000}, where random updates have been used for reasons of limitations on computational resources).

\subsection{Density profiles in the smooth phase}

\begin{figure}[htb]
\begin{center}
\begin{minipage}{0.4\hsize}
\includegraphics[width=\hsize]{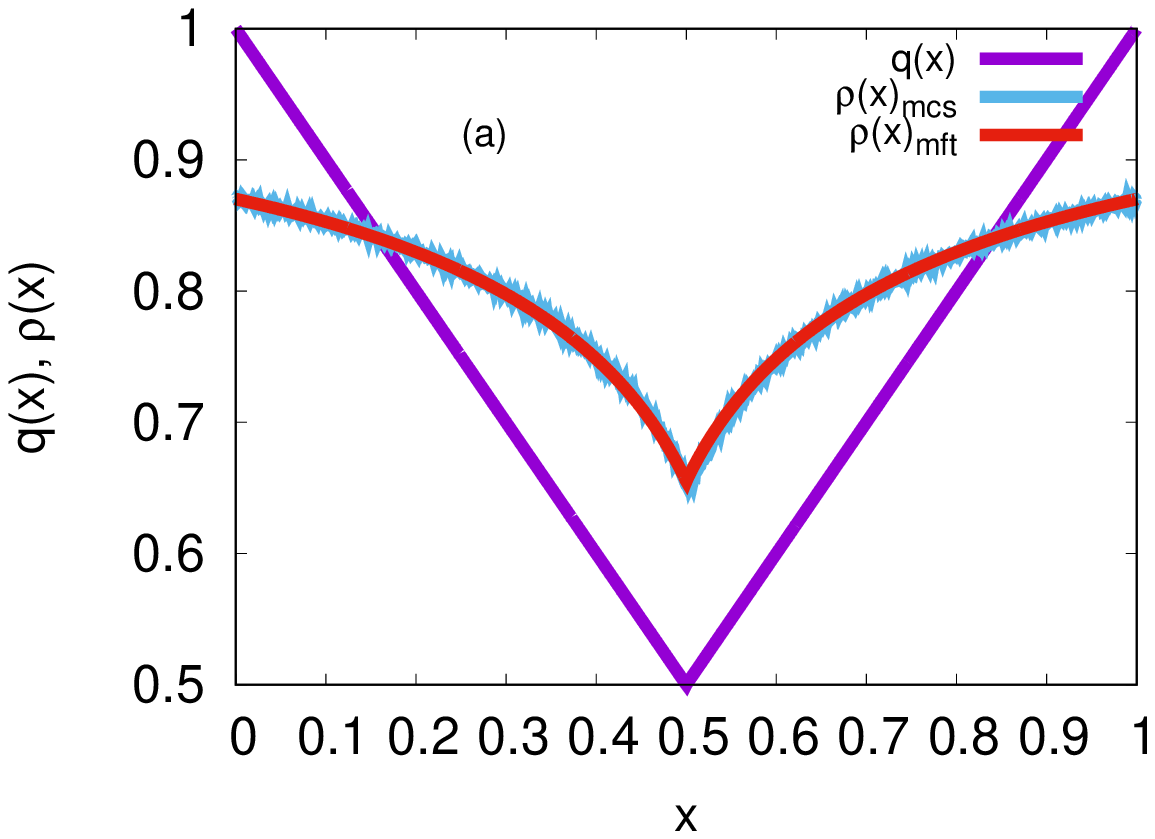}
\end{minipage}
\begin{minipage}{0.4\hsize}
\includegraphics[width=\hsize]{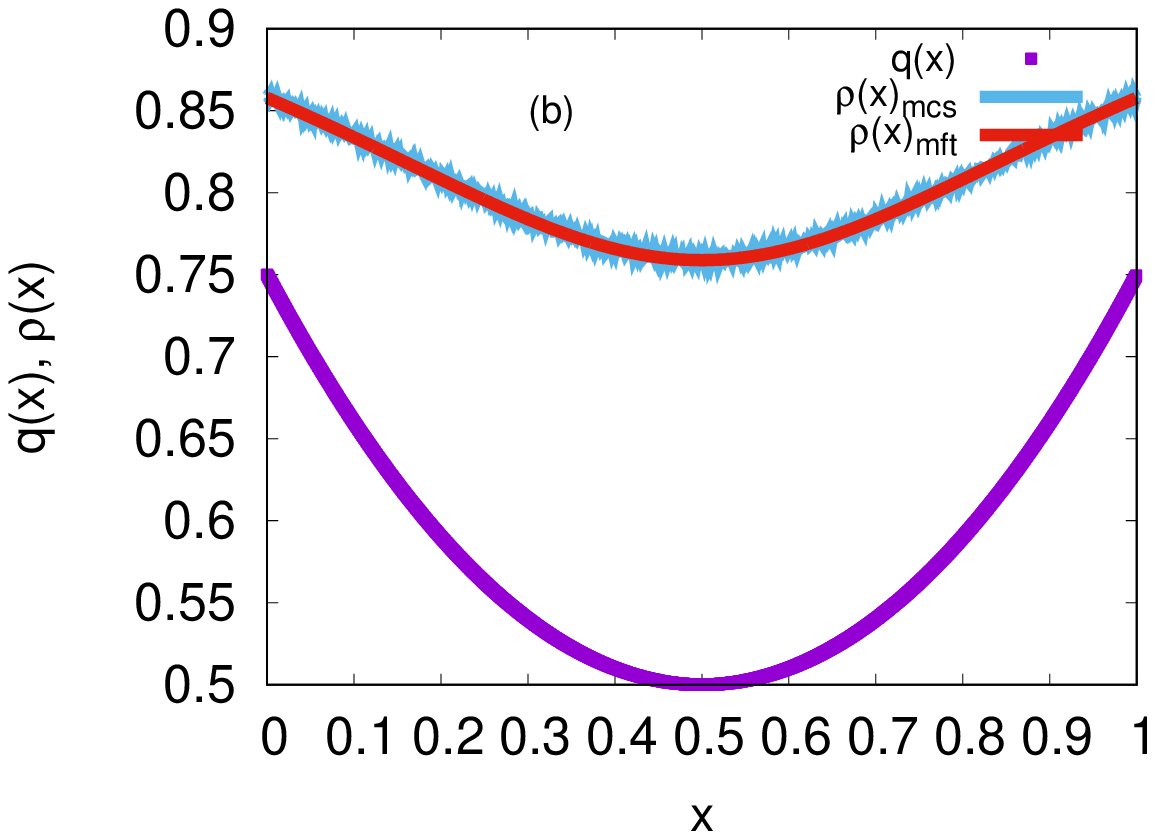}
\end{minipage}
\begin{minipage}{0.4\hsize}
\includegraphics[width=\hsize]{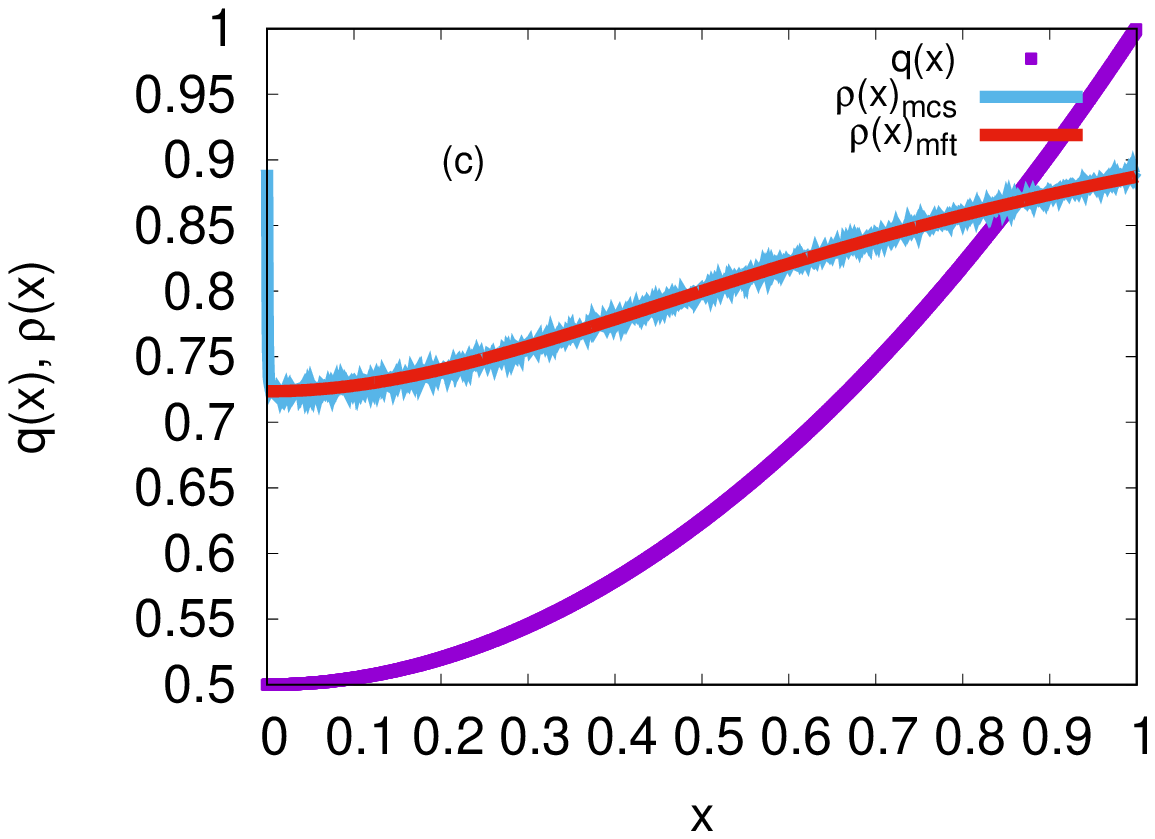}
\end{minipage}
\end{center}
\caption{(a) Plot of $\rho(x)$ versus $x$ in the smooth phase for $q(x)=1-x, 0\leq x<0.5; q(x)=x, 0.5\leq x<1$ (purple continuous line), $n=0.8$. (b) Plot of $\rho(x)$ versus $x$ in the smooth phase for $q(x)=(x-0.5)^2+0.5$ (purple continuous line), $n=0.8$. (c) Plot of $\rho(x)$ versus $x$ in the smooth phase for $q(x)=0.5x^2+0.5$ (purple continuous line), $n=0.8$. Continuous magenta line and overlapping blue points represent, respectively, MFT and MCS data in each plot.}\label{hd-club-app}
\end{figure}

\begin{figure}[htb]
\begin{center}
\begin{minipage}{0.4\hsize}
\includegraphics[width=\hsize]{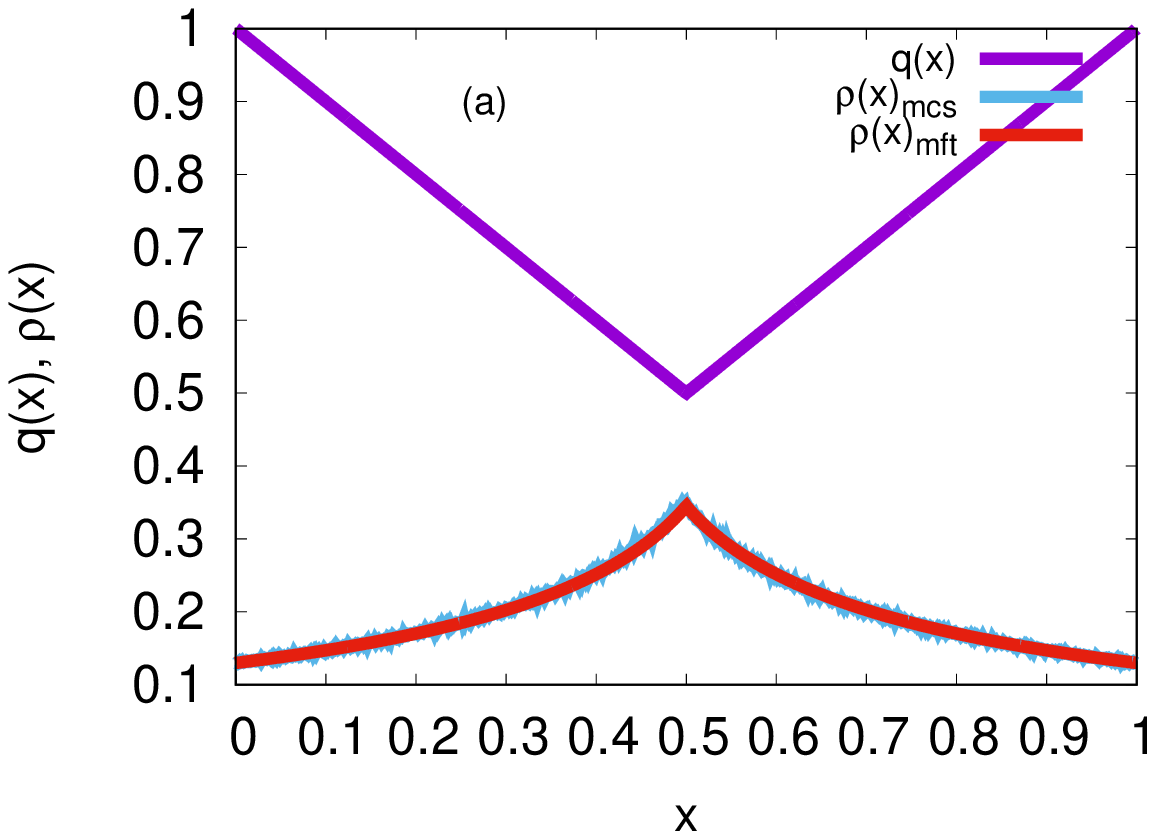}
\end{minipage}
\begin{minipage}{0.4\hsize}
\includegraphics[width=\hsize]{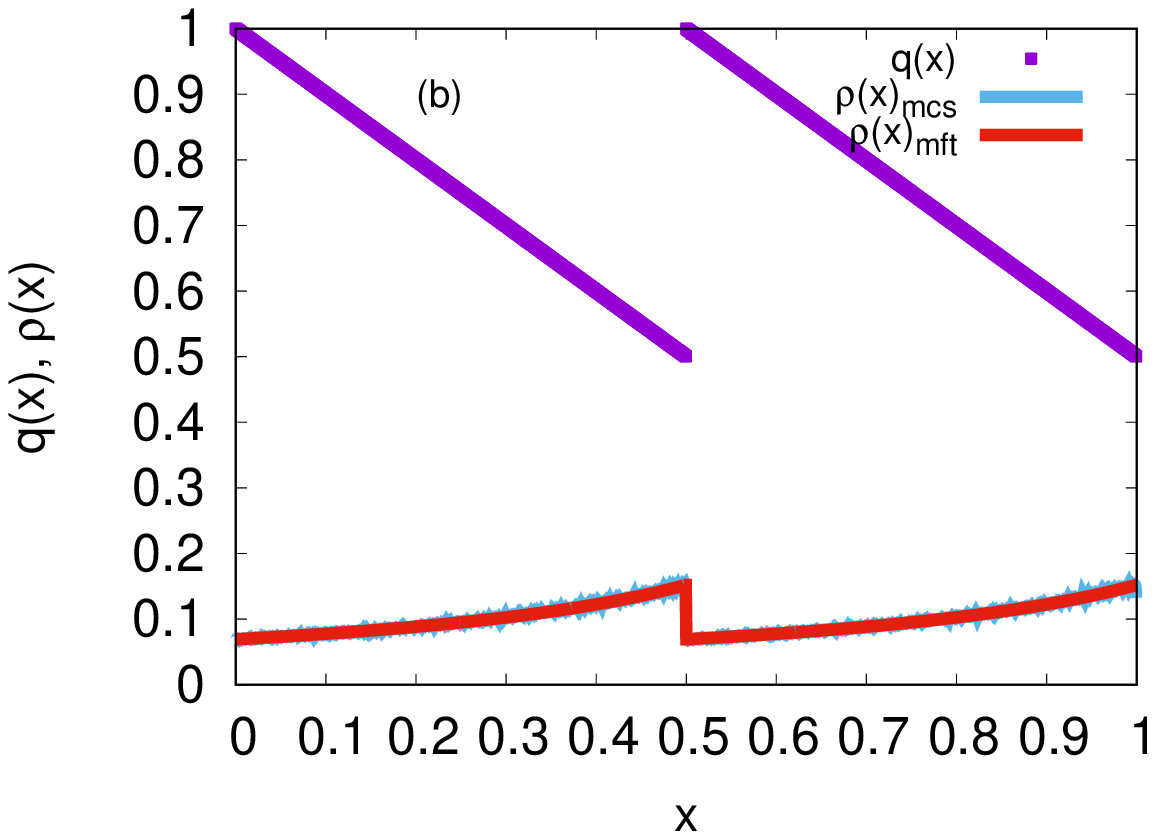}
\end{minipage}
\end{center}
\caption{ (a) Plot of $\rho(x)$ versus $x$ in the smooth phase for $q(x)=1-x, 0\leq x<0.5; q(x)=x, 0.5\leq x<1$ (purple continuous line), $n=0.2$. (b) Plot of $\rho(x)$ versus $x$ in the smooth phase for $q(x)=1-x, 0\leq x<0.5; q(x)=1.5-x, 0.5\leq x<1$ (purple), $n=0.1$ . Continuous magenta line and overlapping blue points represent, respectively, MFT and MCS data in each plot .}\label{ld-club-app}
\end{figure}

\begin{figure}[htb]
\includegraphics[width=8.3cm]{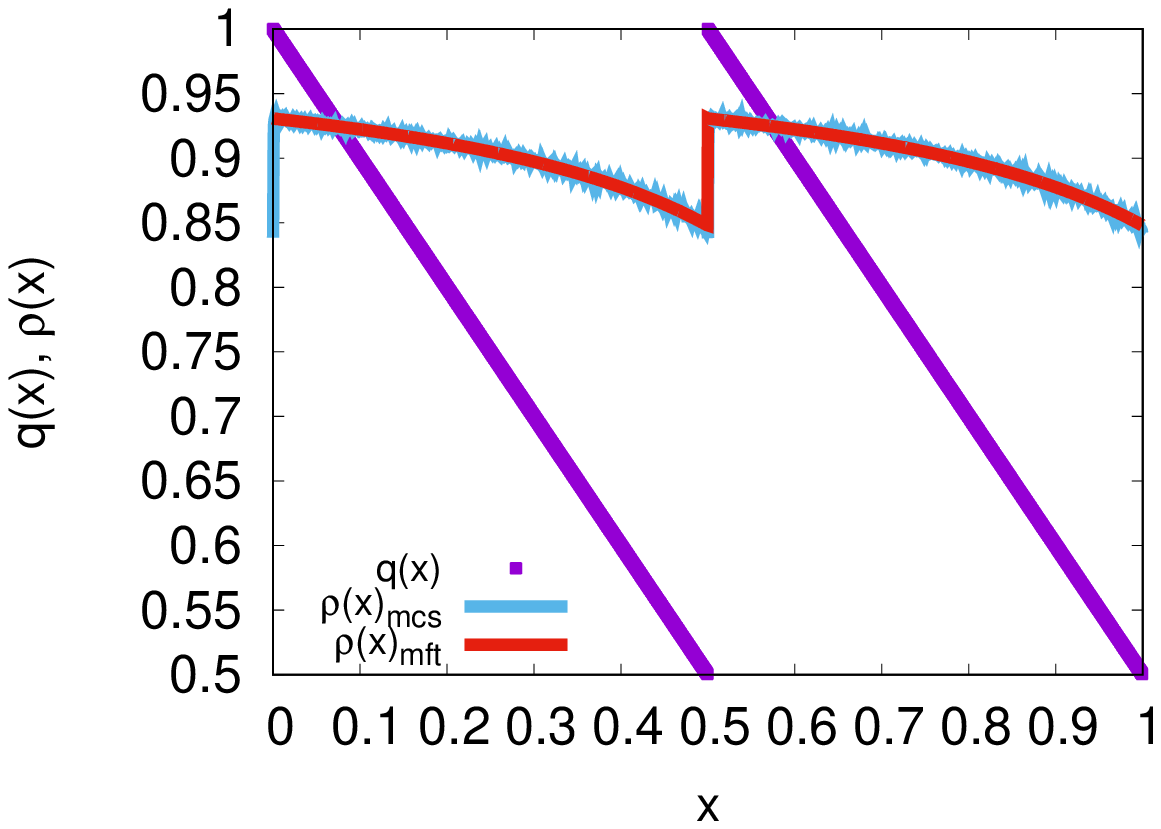}\hfill
\caption{Plot of $\rho(x)$ versus $x$ in the smooth phase for $q(x)=1-x, 0\leq x<0.5; q(x)=1.5-x, 0.5\leq x<1$ (purple), $n=0.9$. Continuous magenta line and overlapping blue points represent, respectively, MFT and MCS data. }  \label{ap6}
\end{figure}

Here we show plots of $\rho(x)$ versus $x$ in the smooth phase with various choices for $q(x)$; see Figs.~(\ref{hd-club-app}-\ref{ap6}).

\subsection{Density profiles in the shock phase}



\begin{figure}[htb]
\begin{center}
\begin{minipage}{0.4\hsize}
\includegraphics[width=\hsize]{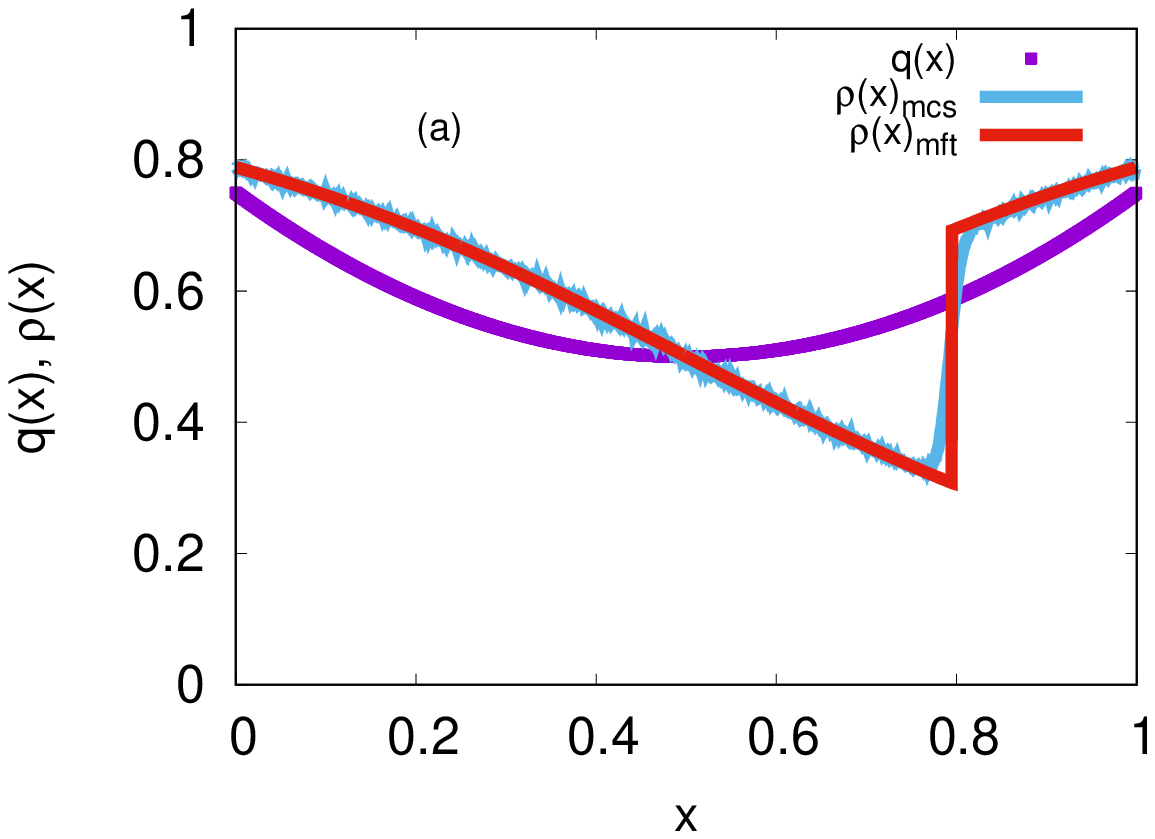}
\end{minipage}
\begin{minipage}{0.4\hsize}
\includegraphics[width=\hsize]{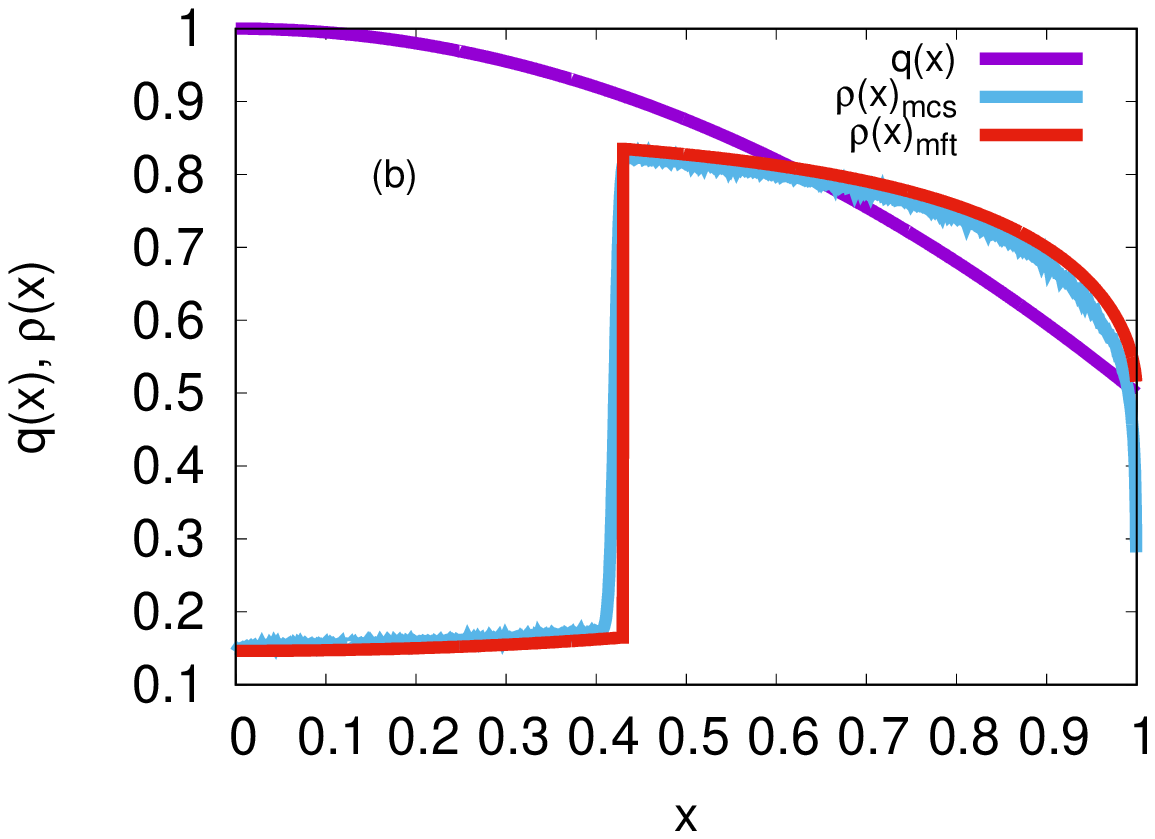}
\end{minipage}
\end{center}
\caption{ (a) Plot of $\rho(x)$ versus $x$ in the shock phase for $q(x)=(x-0.5)^2+0.5$ (purple continuous line), $n=0.6$. (b) Plot of $\rho(x)$ versus $x$ in the shock phase for $q(x)=1-0.5x^2$ (purple continuous line), $n=0.5$. Continuous magenta line and overlapping blue points represent, respectively, MFT and MCS data in each plot .}\label{ldw-club-app}
\end{figure}

{ Here we present a few illustrative examples of the steady state density profiles in the shock phase; see Fig.~\ref{ldw-club-app}.}

\subsection{Density profile in the shock phase with four DDWs}

\begin{figure}[htb]
\begin{center}
\begin{minipage}{0.4\hsize}
\includegraphics[width=\hsize]{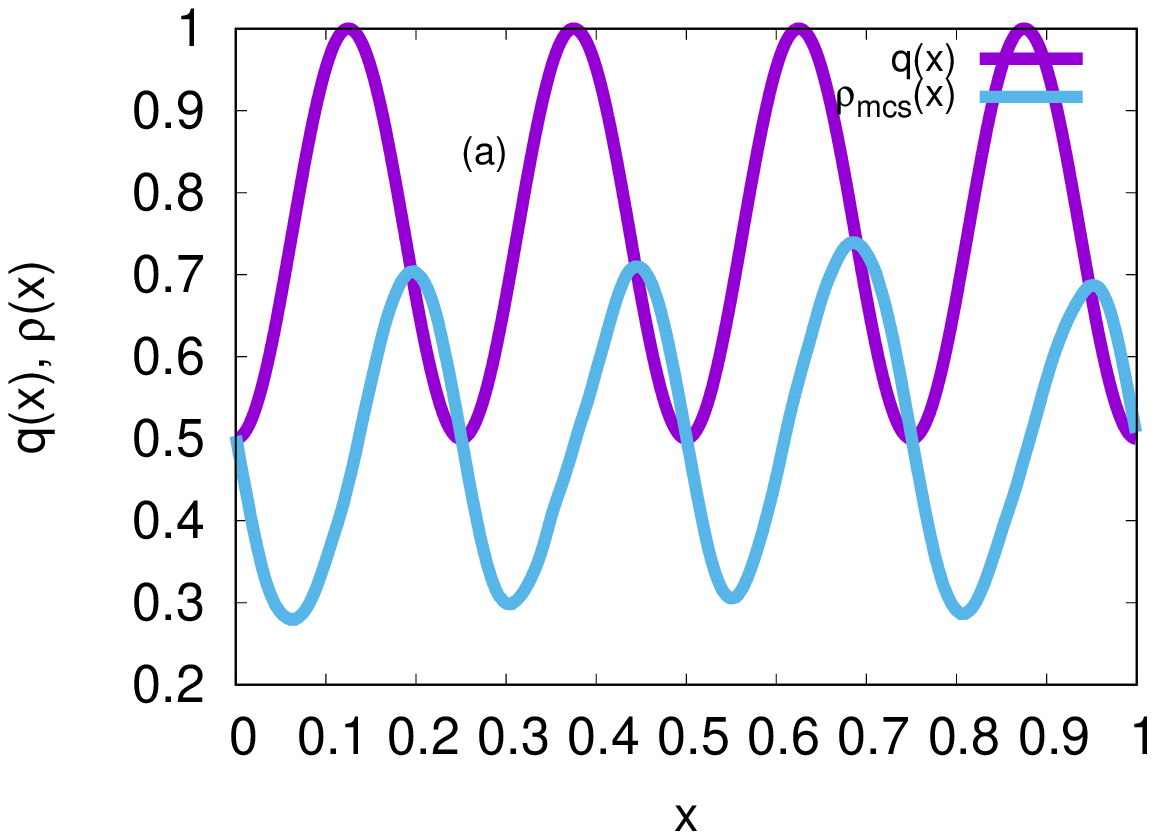}
\end{minipage}
\begin{minipage}{0.4\hsize}
\includegraphics[width=\hsize]{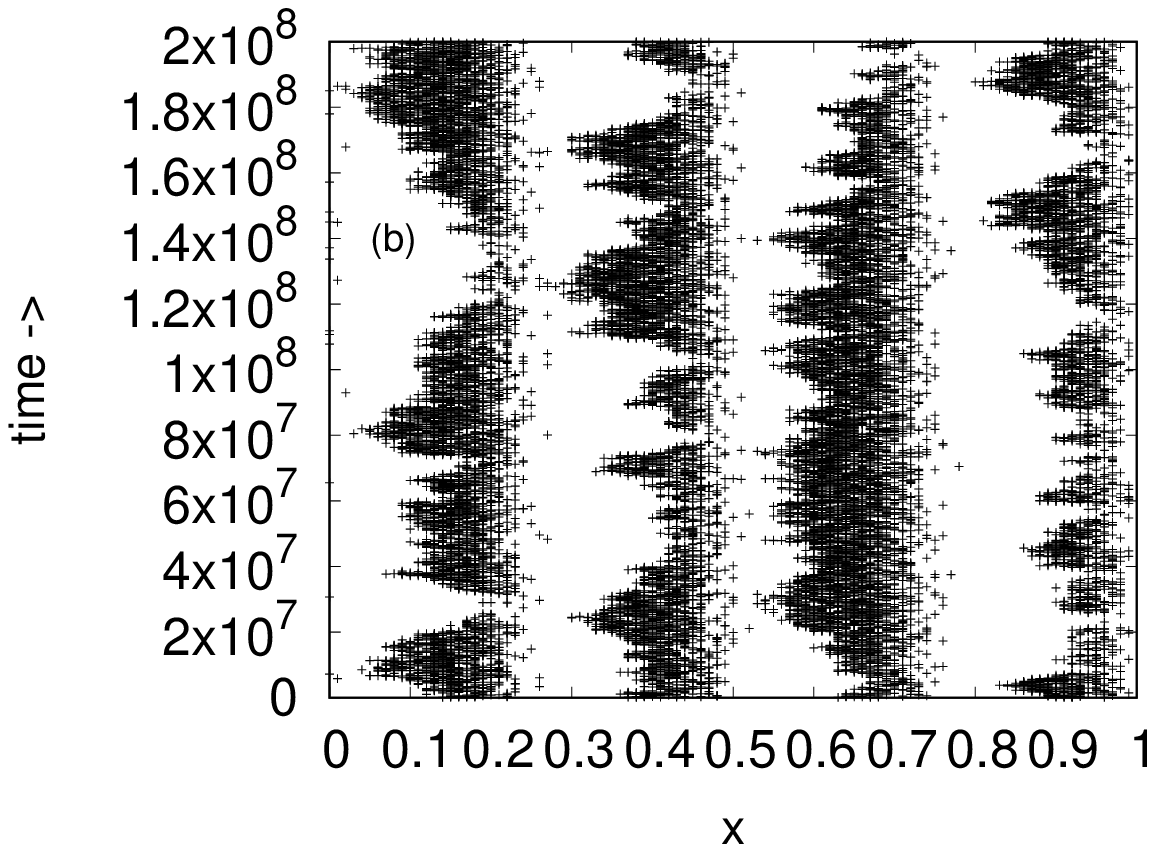}
\end{minipage}
\end{center}
\caption{ (a) MCS plot (blue dashed lines) of $\rho(x)$ versus $x$ for $q(x)=1-0.5\cos^2(4\pi x), n=0.5, N=1200$. The red solid 
lines and the blue dashed lines represent $q(x)$ and MCS data for $\rho(x)$ respectively. (b) Kymograph for $q(x)=1-0.5\cos^2(4\pi x), n=0.5, N=600$. Existence of four DDWs is clearly visible which move without any synchronization between any two of them.}\label{ddw-club-app}
\end{figure}

Here we show results for the density profile when there are four DDWs, see Fig.~\ref{ddw-club-app}.

\begin{thebibliography}{99}



\bibitem{traffic} J. Krug and P. A. Ferrari, {\em J. Phys. A: Math. Gen.} {\bf 29} L465 (1996); D. Chowdhury, L. Santen  and A. Schadschneider, {\em  Phys. Rep. }
{\bf 329},
199 (2000); D. Helbing, {\em Rev. Mod. Phys.} {\bf 73}, 1067 (2001).

\bibitem{traffic2} C. Richer and S. Hasiak, {\em Town Planning Review}, Liverpool University
Press, 2014, 85 (2), pp.217-236.

\bibitem{qdot} T. Karzig and F. von Oppen, {\em Phys. Rev. B} {\bf 81}, 045317 
(2010).

\bibitem{zia1} T Chou, K Mallick, and R K P Zia
{\em Rep. Prog. Phys.} {\bf 74} 116601 (2011).


\bibitem{ribo} 
 S. E. Wells, E. Hillner, R. D. Vale and A. B. Sachs,
{\em Mol. Cell.} {\bf 2}, 135 (1998); S. Wang, K. S. Browning and W.
A. Miller, {\em EMBO J.} {\bf 16}, 4107 (1997). 
\bibitem{pause1} 
Z. A. Afonina {\em et al}, {\em Nucleic Acids Res.} {\bf 42}, 9461 (2014); D. W. Rogers {\em et al}, {\em PLoS Comput. Biol.} {\bf 13} e1005592 (2017).
\bibitem{zia} L Jonathan Cook and R K P Zia J.
Stat. Mech. P02012 (2009); L. Jonathan Cook, R. K. P. Zia, and B.
Schmittmann, Phys. Rev. E 80, 031142 (2009). 



\bibitem{chou1} T. Chou, {\em Biophys. J} {\bf 85}, 755 (2003).



\bibitem{derrida} B. Derrida, E. Domany, D. Mukamel, {\em J. Stat. Phys.} {\bf 69}, 667 (1992); B. Derrida,
S. A. Janowsky, J. L. Lebowitz, E. R. Speer, {\em J. Stat. Phys.} {\bf 78}, 813 (1993); B. Derrida, M.R. Evans, {\em J.
Physique I} {\bf 3}, 311 (1993).


\bibitem{krug-ori} J. Krug, {\em Phys.
Rev. Lett.} {\bf 67}, 1882 (1991).

\bibitem{tasep} R.K.P. Zia, J.J. Dong, and
B. Schmittmann, {\em J. Stat. Phys.} {\bf 144}, 405 (2011).

 \bibitem{stinchcombe} R. B. Stinchcombe and S. L. A. de Queiroz,
{\em Phys. Rev. E} {\bf 83}, 061113 (2011).

\bibitem{mustansir} G. Tripathy and M. Barma, {\em Phys. Rev. E} {\bf
58}, 1911 (1998).

\bibitem{others} M. Bengrine, A. Benyoussef, H. Ez-Zahraouy, and F. Mhirech, {\em Phys. Lett. A} {\bf 253}, 135 (1999); C. Enaud and B. Derrida, {\em Europhys. Lett.} {\bf 66}, 83 (2004).
\bibitem{shaw} L. B. Shaw, J. P. Sethna, and K. H. Lee, {\em Phys. Rev. E} {\bf 70},
021901 (2004).



\bibitem{erwin-rev} A. Parmeggiani, T. Franosch, and E. Frey, {\em Phys. Rev.
Lett.} {\bf 90}, 086601 (2003).

\bibitem{krug-brazi} J. Krug, {\em Brazilian J. of Phys.} {\bf 30}, 97 (2000).

\bibitem{lebo} S. A. Janowsky and J. L. Lebowitz, Phys. Rev. A {\bf 45}, 618 
(1992).

\bibitem{niladri1} N. Sarkar and A. Basu, {\em Phys. Rev. E} {\bf 90}, 022109 (2014).

\bibitem{frey}  P. Pierobon, M. Mobilia, R. Kouyos and E. Frey, {\em Phys. Rev. E} {\bf 74}, 031906 (2006).



\bibitem{tirtha1} T. Banerjee, N. Sarkar and A. Basu {\em JStat. Mech.: Theory 
and Experiment}
  P01024 (2015).

\bibitem{stinchcombe1} R. J. Harris and R. B. Stinchcombe, {\em Phys. Rev. E} {\bf 70}, 016108 (2004).


\bibitem{dna} B Alberts, A Johnson, J Lewis, M Raff, K Roberts, and P Walter, {\em Molecular
Biology of the Cell} Garland Science, 4th edition (2002); B Li, M Carey, and J L Workman, {\em
Cell} {\bf 128}, 707 (2007); C Y Lin, J Lov\'en, P B Rahl, R M Paranal, C B Burge, J E Bradner, T I
Lee, and R A Young, {\em Cell} {\bf 151}, 56 (2012); J L Workman and R E Kingston, {\em Annual Review of Biochemistry} {\bf 67}, 545 (1998).
\bibitem{dan} D. D. Erdmann-Pham, K. D. Duc and Y. S. Song, {\em arXiv: 1803.05609}.

\bibitem{bechinger-jphys} I. Buttinoni {\em et al}, {\em J. Phys. Condens. Matter} {\bf 24}, 284129 (2012).

\bibitem{bechinger} Q.-H. Wei, C. Bechinger, and P. Leiderer, Science 287,
625 (2000); C. Lutz, M. Kollmann, and C. Bechinger, Phys. Rev. Lett. {\bf 93}, 026001 (2004).

\bibitem{bechinger-rmp} C. Bechinger {\em et al}, {\em Rev. Mod. Phys.} {\bf 88}, 045006 (2016).

\bibitem{erwin-tobias} T. Reichenbach, T. Franosch, and E. Frey, {\em Eur. Phys. J E} {\bf 27}, 47 (2008).
\bibitem{kramers} U. T\"auber, {\em Critical Dynamics}, Cambridge University Press (Cambridge, 2014).

\end{thebibliography}
\end{document}